\documentclass[amsmath, amsfonts, superscriptaddress, showpacs, prb, twocolumn]{revtex4}
\usepackage{graphicx}
\usepackage{epsfig}
\usepackage{bm}
\usepackage{dcolumn}
\usepackage{amsmath}
\usepackage{amssymb}

\begin{document}

\title{Interaction-induced corrections to conductance and
thermopower in quantum wires}

\author{Alex Levchenko}
\affiliation{Materials Science Division, Argonne National
Laboratory, Argonne, Illinois 60439, USA}

\author{Zoran Ristivojevic}
\affiliation{Materials Science Division, Argonne National
Laboratory, Argonne, Illinois 60439, USA}

\affiliation{CNRS-Laboratoire de Physique Th\'{e}orique de l'Ecole
Normale Sup\'{e}rieure, 73231 Paris Cedex 05, France}

\author{Tobias Micklitz}
\affiliation{Dahlem Center for Complex Quantum Systems and Institut
f\"{u}r Theoretische Physik, Freie Universit\"{a}t Berlin, 14195
Berlin, Germany}

\begin{abstract}
We study transport properties of weakly interacting spinless
electrons in one-dimensional single channel quantum wires. The
effects of interaction manifest as three-particle collisions due to
the severe constraints imposed by the conservation laws on the
two-body processes. We focus on short wires where the effects of
equilibration on the distribution function can be neglected and
collision integral can be treated in perturbation theory. We find
that interaction-induced corrections to conductance and thermopower
rely on the scattering processes that change number of right- and
left-moving electrons. The latter requires transition at the bottom
of the band which is exponentially suppressed at low temperatures.
Our theory is based on the scattering approach that is beyond the
Luttinger-liquid limit. We emphasize the crucial role of the
exchange terms in the three-particle scattering amplitude that was
not discussed in the previous studies.
\end{abstract}

\date{December 26, 2010}


\maketitle

\textit{Introduction}.-- The quest for fundamental theory of
interacting low-dimensional many-body quantum liquids and solids
continues.~\cite{Khodas,Lunde,Pereira,JTK,IG,Sirker,Review,TJK,MAP,ATJK}
Over the past several decades the traditional framework for
discussing one-dimensional (1D) systems was provided by the
Luttinger-liquid (LL) theory.~\cite{LL} It exploits an approximation
of linearized fermionic dispersion relation which makes this model
exactly solvable. While being extremely fruitful in many cases an
ideal LL model, however, possesses certain deficiency. For example,
elementary bosonic excitations (plasmons) of the LL have infinite
life time so that there is no relaxation toward the equilibrium
within this description regardless of the strength of interaction.
The effects of interaction show no sign in application to the dc
transport coefficients of clean, single channel short quantum wires.
Indeed, it has been shown~\cite{MS-theorem} that within LL model the
interactions inside the wire do not affect conductance which is
simply given by its noninteracting value. Clearly, a model with the
linearized spectrum is an idealized one. It is really the delicate
interplay between the dispersion nonlinearity and interactions in
reduced dimensions that bring new insights. Thus, current interest
in the properties of 1D systems is focused on the physics that lies
beyond the Luttinger-liquid limit.

In support of this general interest a number of recent experiments
in the low-density quantum wires revealed deviations from the
perfect conductance quantization,~\cite{G-exp} a lower value of the
thermal conductance than predicted by the Wiedemann-Franz
law,~\cite{K-Exp} and enhanced thermopower.~\cite{S-Exp} Although
there is no consensus on the theoretical interpretation of these
observations it is widely accepted that interaction effects are
crucial in understanding of these features. The goal of this paper
is to elucidate the role of inelastic scattering processes (i.e.
processes which change the number and energy of, say, right-moving
electrons), not captured by the LL theory, for the description of
transport properties of one-dimensional quantum wires. We evaluate
the interaction-induced corrections to conductance and thermopower
in a particularly interesting case of spinless (spin-polarized)
electrons. Unlike the recent work, see Ref.~\onlinecite{Lunde},
where the model of interacting electrons with sharp momentum cut-off
was used, we consider more generic situation which requires to
account for both the direct and \textit{exchange} terms in the
three-particle scattering rate.

\begin{figure}[b!]
  \includegraphics[width=8cm]{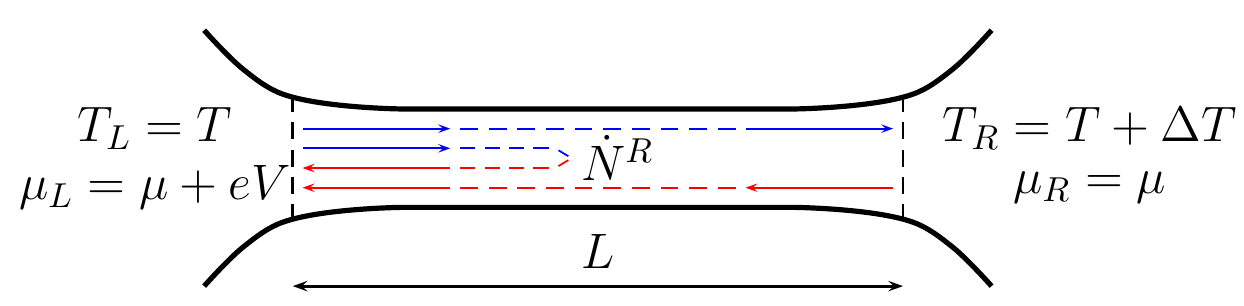}\\
  \caption{[Color online] Quantum wire
of length $L$ in the regime of small voltage $\mu_L-\mu_R=eV$ and/or
temperature $T_R-T_L=\Delta T$ bias. The interaction-induced
backscattering changes the number of right-moving electrons,
$\dot{N}^R\neq0$, thus affecting electrical conductance $G$ and
thermopower $S$.}\label{Fig1}
\end{figure}

\textit{Formalism}.-- We consider a clean single channel quantum
wire connected adiabatically to the bulk noninteracting leads, and
biased by a small voltage $V$ and temperature difference $\Delta T$,
see Fig.~\ref{Fig1}. The distribution function of noninteracting
electrons is purely determined by the leads and has the form
\begin{align}\label{f}
f_p=\frac{\theta(p)}{e^{\frac{\varepsilon_p-\mu_L}{T_L}}+1}
+\frac{\theta(-p)}{e^{\frac{\varepsilon_p-\mu_R}{T_R}}+1}
\end{align}
where the chemical potentials and temperatures in the leads are
$\mu_L=\mu+eV$, $T_L=T$ and $\mu_R=\mu$, $T_R=T+\Delta T$. The
energy of an electron with momentum $p$ is $\varepsilon_p=p^2/2m$,
and $\theta(p)$ is the unit step function.

In the general interacting case even weak processes of electron
scattering will modify nonequilibrium distribution Eq.~\eqref{f}.
Indeed, some right-moving electrons will \textit{backscatter} thus
become left-movers. In that way electrons lose the memory of the
lead the originated from and tend to equilibrate. To what extent the
equilibration occurs depends on the strength of interaction and
length of the wire.~\cite{TJK}

Relying only on the very general basis of particle number
conservation, one may show that the electric current $I$ flowing
through the wire is ultimately related to the electron
backscattering~\cite{Lunde,JTK}
\begin{equation}\label{1}
GV=I-e\dot{N}^R
\end{equation}
where $\dot{N}^R$ is the rate of change in the number of
right-moving electrons, see Fig.~\ref{Fig1}. The physical meaning of
Eq.~\eqref{1} is clear: without interaction $\dot{N}^R=0$ and one
then finds the Landauer conductance of noninteracting electrons,
$G=I/V=(e^2/h)(1+e^{-\mu/T})^{-1}$, which coincides with the
conductance quantum, $e^2/h$, up to an exponentially small
correction, $\sim e^{-\mu/T}$, due to the states at the bottom of
the band. In the presence of interaction some right-moving electrons
are backscattered reducing the current. Therefore, the
interaction-induced correction to conductance is $\delta
G\propto\dot{N}^R$ and Eq.~\eqref{1} can be considered as the
generalization of Landauer formula.

Apart from the conductance we are interested also in the
thermopower, $S$, which relates induced voltage across the wire to
applied temperature difference. For the noninteracting electrons
$S=V/\Delta T|_{I=0}\simeq(\mu/eT)e^{-\mu/T}$, to the leading order
at low temperatures $T/\mu\ll1$. Within the LL model $S=0$ due to
particle-hole symmetry. The reason for such strong suppression of
thermopower is the partial cancelation between heat currents carried
by electron-like and hole-like excitations. Only the absence of
electronic states below the bottom of the band prevents thermopower
from vanishing exactly. Knowing $S$ one can also find the Peltier
coefficient $\Pi$ via the Onsager relation $\Pi=ST$. As we discuss
below the interaction-induced corrections to thermopower also result
from the electron backscattering, so that $\delta
S\propto\dot{N}^R$.

Within the Luttinger-liquid model we have $\dot{N}^R=0$. It is
because the constraints imposed by the energy and momentum
conservations allow either zero-momentum exchange or an interchange
of the momenta for two colliding particles. In either case $f_p$
given by Eq.~\eqref{f} remains unchanged. Thus, the transport
coefficients $G$ and $S$ remain intact by two-body interactions.
Therefore, one can conclude then that the leading backscattering
mechanism is due to three-particle
collisions.~\cite{Lunde,JTK,TJK,ATJK}

\begin{figure}
  \includegraphics[width=8cm]{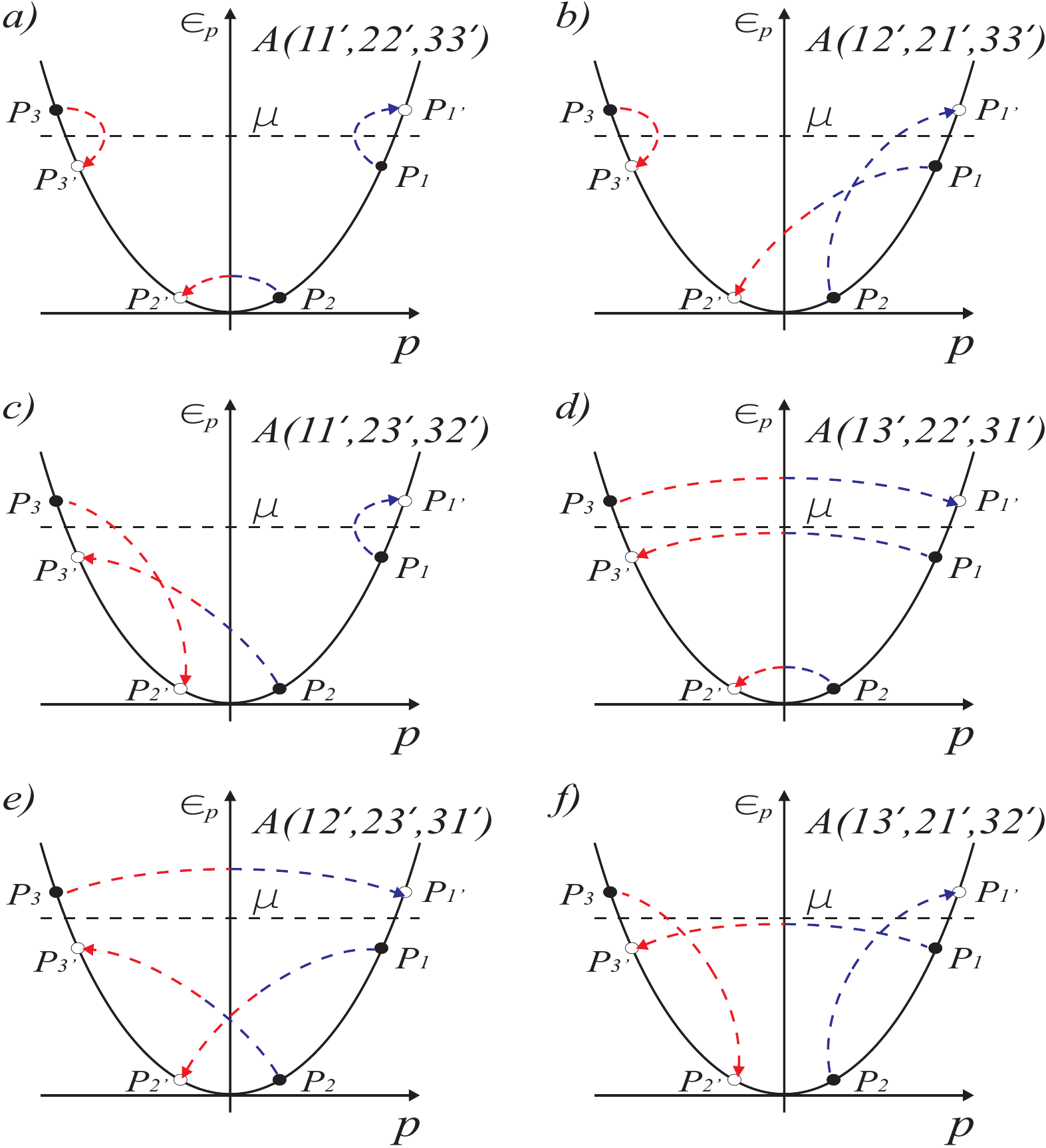}\\
  \caption{[Color online] Three-particle collision processes that sum up into the
  full amplitude in Eq.~\eqref{A} for a scattering which involves one
  particle at the bottom of the band and the other two near the
  opposite Fermi points. The uppermost left figure-(a) represents
  the direct term in the scattering amplitude while the other five
  (b)--(f) are the exchange contributions. These are the dominant
  three-particle processes that contribute to
  $\dot{N}^R$ and
  thus renormalize conductance and  thermopower.} \label{Fig2}
\end{figure}

Rather than working within the LL model with nonlinear dispersion,
that would include the anharmonic interaction of plasmons and thus,
in principle, contain equilibration and backscattering processes, we
consider simpler situation of weakly interacting electrons,
$e^2/\hbar v_F\kappa\lesssim1$, where $v_F$ is Fermi velocity while
$\kappa$ dielectric constant. We account for the three-particle
collisions within the Boltzmann equation (BE) formalism. The
evolution of the distribution function of an interacting many-body
system is governed by the BE, $\dot{f}_{p}=\mathcal{I}\{f_{p}\}$,
where the collision integral is given by
\begin{eqnarray}\label{I-coll}
&&\hskip-.7cm\mathcal{I}\{f_{p_1}\}=-\hskip-.3cm\sum_{p_2p_3\atop
p_{1'}p_{2'}p_{3'}}\hskip-.3cm
W^{1'2'3'}_{123}[f_{p_1}f_{p_2}f_{p_3}(1-f_{p_{1'}})(1-f_{p_{2'}}) \nonumber\\
&&\hskip-.7cm\times(1-f_{p_{3'}})-
f_{p_{1'}}f_{p_{2'}}f_{p_{3'}}(1-f_{p_1})(1-f_{p_2})(1-f_{p_3})]
\end{eqnarray}
In general collision integral is a nonlinear functional of $f_p$
which we assumed here to be local in space. A particular combination
of the distribution functions in Eq.~\eqref{I-coll} conventionally
accounts for the probability to find filled $f_{p_i}$ and empty
$1-f_{p_{i'}}$ states before and after the collision. The key
element of the theory is the scattering rate
$W^{1'2'3'}_{123}=(2\pi/\hbar)|\mathcal{A}^{1'2'3'}_{123}|^2\delta(E-E')\delta_{P,P'}$
from the set of initial states $\{p_1,p_2,p_3\}$ into the final
states $\{p_{1'},p_{2'},p_{3'}\}$. The delta-functions in
$W^{1'2'3'}_{123}$ impose conservation of total energy
$E(E')=\sum_{i}\varepsilon_{p_{i}(p_{i'})}$ and total momentum
$P(P')=\sum_{i}p_{i(i')}$ in a collision, and
$\mathcal{A}^{1'2'3'}_{123}$ is corresponding scattering amplitude.
For the electrons with the bare pair-interaction potential
 three-particle amplitude can be found via generalized
Fermi golden rule by iterating interaction to the second order. The
result of such calculation gives~\cite{Lunde}
\begin{equation}\label{A}
\mathcal{A}^{1'2'3'}_{123}=\sum_{\pi(1'2'3')}\mathrm{sign}(1'2'3')A(11',22',33')
\end{equation}
where $A(11',22',33')$ is an amplitude of a particular scattering
process, see Fig.~\ref{Fig2}, which reads explicitly
\begin{eqnarray}\label{A-aaa}
&&\hskip-.55cm
A(11',22',33')=a^{1'2'}_{12}+a^{1'3'}_{13}+a^{2'3'}_{23}\\
\label{A-123} &&\hskip-.55cm a^{1'2'}_{12}\equiv
a^{p_{1'}p_{2'}}_{p_1p_2}=\frac{1}{L^2}V_{p_{1'}-p_1}V_{p_{2'}-p_2}\times\\
&&\hskip-.55cm
\left[\frac{1}{E\!-\!\varepsilon_{p_1}\!-\!\varepsilon_{p_{2'}}\!-\!\varepsilon_{P-p_2-p_{1'}}}
+\frac{1}{E\!-\!\varepsilon_{p_{1'}}\!-\!\varepsilon_{p_2}\!-\!\varepsilon_{P-p_{2'}-p_1}}\right]\nonumber
\end{eqnarray}
Here $\pi(\ldots)$ and $\mathrm{sign}(\ldots)$ denote permutations
of the final momenta and parity of a particular permutation, finally
$V_p$ is the Fourier component of the bare two-body interaction
potential.

In the following we calculate $\dot{N}^R$ from the BE, having in
mind that the distribution function at the ends of the wire are
Fermi functions determined by the leads, Eq.~\eqref{f}. The rate of
change in the number of right-moving electrons is defined as
$\dot{N}^R=\sum_{p>0}\dot{f}_p=\sum_{p>0}\mathcal{I}\{f_p\}$. Since
we restrict our analysis to very short wires in which three-particle
collisions are rare and thus effect of relaxation on the
distribution function can be neglected, we can treat collision
integral [Eq.~\eqref{I-coll}] in perturbation theory. Expanding the
distribution function in Eq.~\eqref{f} to the linear order in $V$
and $\Delta T$ as $f_p\simeq
f^0_p+f^0_p(1-f^0_p)\left[\frac{eV}{T}\theta(p)+\frac{(\varepsilon_p-\mu)\Delta
T}{T^2}\theta(-p)\right]$, where
$f^0_p=[e^{(\varepsilon_p-\mu)/T}+1]^{-1}$ is equilibrium Fermi
function, we get with the help of Eq.~\eqref{I-coll}
\begin{eqnarray}\label{NR}
\dot{N}^R\!=3\!\!\sum_{++-\atop
+--}\!\mathcal{W}^{1'2'3'}_{123}\!\!\left[\frac{\Delta
T}{T^2}(\varepsilon_{p_{3'}}-\varepsilon_{p_3}+\varepsilon_{p_{2'}}-\mu)-\frac{eV}{T}\right]
\end{eqnarray}
where
$\mathcal{W}^{1'2'3'}_{123}\!\!=\!W^{1'2'3'}_{123}f^0_{p_1}f^0_{p_2}f^0_{p_3}(1\!-\!f^0_{p_{1'}})
(1\!-\!f^0_{p_{2'}})(1\!-\!f^0_{p_{3'}})$. The notation
$\sum_{++-\atop +--}$ means $\sum_{p_1>0,p_2>0,p_3<0\atop
p_{1'}>0,p_{2'}<0,p_{3'}<0}$, etc. In deriving Eq.~\eqref{NR} we
have used the symmetry of the scattering rate under: (i) the
interchange of all primed and unprimed indices; (ii) a pairwise
exchange and (iii) the inversion of all momenta $p_i\to-p_i$. In the
course of the derivation one can see that in order to have
$\dot{N}^R\neq0$ the number of right-moving electrons must change
after the collision. In Eq.~\eqref{NR} we have kept only the leading
contribution to $\dot{N}^R$, namely an electron backscattered by one
right mover and one left mover which preserve their direction of
motion after the collision. This particular scattering process is
compatible with the conservation laws, it involves only a single
state at the bottom of the band, and it has all final scattering
states in the vicinity of the initial ones. At low temperatures we
expect $\dot{N}^R\propto e^{-\mu/T}$ which stems from the Fermi
occupation factors for the particle at the bottom of the band.

\textit{Scattering amplitudes}.-- Let us look closer at the
kinematics of backscattering processes. In general three-particle
scattering amplitude [Eq.~\eqref{A}] depends on all $p_i$ and
$p_{i'}$. However, for the momentum configuration under
consideration [see Fig.~\ref{Fig2}] $p_{2(2')}$ lies near the bottom
of the band, while $p_{1(1')}$ and $p_{3(3')}$ lie near the right
and left Fermi points, all within a small range $|p_i-p_{i'}|\sim
T/v_F$ set by temperature $T\ll\mu$. We thus argue that, up to small
corrections in $\delta p\sim T/v_F\ll p_F$, one can replace
$p_1\simeq+p_F$, $p_2\simeq0$ and $p_3\simeq-p_F$ in the expressions
of Eqs.~\eqref{A}--\eqref{A-123}, which then becomes a function of
$q_i=p_{i'}-p_i$ only
$\mathcal{A}^{1'2'3'}_{123}\to\mathcal{A}(q_1,q_2,q_3)$, with $q_i$
being momenta transferred in a collision. Furthermore, using the
approximate forms of dispersion relation near Fermi points
$\varepsilon_{p_{1'}}-\varepsilon_{p_1}\approx v_Fq_1$ and
$\varepsilon_{p_{3'}}-\varepsilon_{p_3}\approx-v_Fq_3$ the
conservation laws allow one to express $q_1$ and $q_3$ in terms of
$p_2$ and $p_{2'}$ as
$q_{1,3}=\frac{p_2-p_{2'}}{2}\mp\frac{\varepsilon_{p_{2'}}-\varepsilon_{p_2}}{2v_F}$.
One readily sees that $q_1\simeq q_3\simeq-q_2/2$ up to small
contributions of order $p_2/p_F\ll1$, such that amplitude
effectively depends on a single momentum,
$\mathcal{A}(q_2)\equiv\mathcal{A}(-q_2/2,q_2,-q_2/2)$. Applying
these observations to Eq.~\eqref{A} we can expand amplitude for
$|q_2|\sim T/v_F\ll p_F$ to the leading order and obtain
\begin{eqnarray}\label{A-Approx}
\mathcal{A}(q_2)=\frac{1}{2\mu
L^2}\left\{V_{p_F}[V_{q_2/2}-V_{2p_F}]-V_{2p_F}[V_{q_2}-V_{2p_F}]\right.\nonumber\\
-V_{q_2/2}[V_{q_2/2}-V_{q_2}]+2p_FV'_{2p_F}[V_{q_2}-V_{p_F}]\nonumber\\
\left.-2p_FV'_{p_F}[V_{q_2/2}-V_{p_F}]+p_FV'_{p_F}[V_{q_2/2}-V_{2p_F}]\right\}
\end{eqnarray}

There are several useful checks we can make at this point. It is
known from the context of \textit{integrable} quantum many-body
problems~\cite{Sutherland} that for some two-body potentials,
scattering of the particles of $N$-body system factorizes into a
sequence of two-body collisions. In the context of this work, it
means that three-particle scattering for the \textit{integrable}
potentials may result only in permutations within the group of three
momenta of the colliding particles; all other three-particle
scattering amplitudes must be exactly zero for such potentials. We
have checked explicitly that the three-particle scattering amplitude
in Eq.~\eqref{A} nullifies for the several special potentials: for
the contact interaction, $V_p\propto\mathrm{const}$, for the
Calogero-Suthreland model, $V_p\propto|p|$, and for the fermionic
equivalent of the Lieb-Liniger model, $V_p\propto 1-p^2/p^2_0$. Of
course, the simplified version of the amplitude
[Eq.~\eqref{A-Approx}] obtained for a specific scattering process
that we need [Fig.~\ref{Fig2}] also vanishes for these potentials.
For generic \textit{non-integrable} models the three-particle
amplitude is not expected to be zero. In the following we take
regularized Coulomb potential $1/|x|\to 1/\sqrt{x^2+4\Lambda^2}$
where cut-off $\Lambda=d$ is distance to the screening gate at large
$x$ while $\Lambda=w$ is the wire width at small $x$. Thus, the
Fourier transformed component of the interaction reads
$V_p=(2e^2/\kappa)[K_0(2|p|w/\hbar)-K_0(2|p|d/\hbar)]$, where
$K_0(z)$ is Bessel function. In particular, for the case of screened
Coulomb interaction, when $p\ll\hbar/d\ll\hbar/w$, we find from
Eq.~\eqref{A-Approx}
\begin{equation}\label{A-SC}
\mathcal{A}(q_2)=-3(\ln4-1)(2e^2/\kappa)^2\lambda_s(k_Fd)/\mu L^2
\end{equation}
where $\lambda_s(x)=x^4\ln(1/x)$, while for the unscreened case,
when $\hbar/d\ll p\ll\hbar/w$, we obtain
\begin{equation}\label{A-UC}
\mathcal{A}(q_2)=(3/4)(2e^2/\kappa)^2\lambda_u(k_Fw)\ln(p_F/|q_2|)/\mu
L^2
\end{equation}
where $\lambda_u(x)=x^2\ln(1/x)$. Note, these forms of the amplitude
require to keep all the \textit{exchange} terms in Eq.~\eqref{A}.
Unlike in the the previous studies~\cite{Lunde,TJK}, which assumed
interaction with sharp momentum cut-off, keeping only direct term in
the amplitude would give sub-leading contribution, namely
$\mathcal{A}_{\mathrm{dir}}(q_2)\propto q^2_2\ln|q_2|$. Indeed,
comparing this to the exchange contribution in Eq.~\eqref{A-UC} and
using $q_2\sim T/v_F$ one estimates
$|\mathcal{A}_{\mathrm{dir}}|/|\mathcal{A}_{\mathrm{ex}}|\sim(T/\mu)^2\ll1$.

\textit{Results and discussions}.-- Having determined scattering
amplitude we are prepared to compute the rate of change in the
number of right-movers due to electron backscattering. In accordance
with our earlier kinematic observations we approximate conservation
laws implicit in the scattering rate $W^{1'2'3'}_{123}$ of
Eq.~\eqref{NR} as
$\delta(E-E')\delta_{P,P'}\simeq\frac{1}{v_F}\delta(q_1-q_3)\delta_{2q_1+q_2=0}$,
which removes $q_{2,3}$ integrals. We can also complete $p_{1,3}$
integrals exactly
\begin{equation}
\sum_{p}f^0_{p}(1-f^0_{p+q})=\frac{L}{2h}\frac{q}{\sinh\frac{v_Fq}{2T}}e^{\pm
v_Fq/2T}
\end{equation}
for $p$ near $\pm p_F$. This gives all together
\begin{eqnarray}\label{NR-leading}
\hskip-.25cm \dot{N}^R\!=\!-\frac{3L^3}{16\pi^2\hbar^4
v_F}\!\left[\frac{\mu\Delta T}{T^2}\!+\!\frac{eV}{T}\right]\!\!
\sum_{p_2>0,q_1}\!\!\!\!\frac{q^2_1|\mathcal{A}|^2M(p_2,q_1)}{\sinh^2\frac{v_Fq_1}{2T}}
\end{eqnarray}
where $M(p_2,q_1)=\theta(-p_2+2q_1)f^0_{p_2}(1-f^0_{p_{2}-2q_1})$.
Since characteristic $p_2$ lies at the bottom of the band we can
also approximate $f^0_{p_2}(1-f^0_{p_{2}-2q_1})\approx
e^{-\mu/T}\big[1+\frac{\varepsilon_{p_2-2q_1}}{2mT}\big]$ to the
leading order at small temperatures, since
$\frac{\varepsilon_{p_2-2q_1}}{2mT}\sim\frac{(T/v_F)^2}{mT}\sim(T/\mu)\ll1$.
Then, $p_2$ summation gives factor of
$(2Lq_1\theta(q_1)/h)e^{-\mu/T}$ and remaining $q_1$ integration in
Eq.~\eqref{NR-leading} is straightforward. As a result, we find from
$\delta G=e\dot{N}^R/V$ for the case of screened Coulomb interaction
with the amplitude taken from Eq.~\eqref{A-SC}, the
interaction-induced correction to conductance
\begin{eqnarray}\label{G}
\hskip-.2cm\delta
G=-c(e^2/h)(k_FL)r^4_s\lambda^2_s(k_Fd)(T/\mu)^3e^{-\mu/T}
\end{eqnarray}
where $r_s=e^2/\hbar v_F\kappa$ and
$c=\frac{324\zeta(3)}{\pi^3}\ln^2(4/e)$, while from $\delta
S=-h\dot{N}^R/e\Delta T$ correction to thermopower (restoring
Boltzmann constant $k_B$)
\begin{equation}\label{S}
\delta S=c(k_B/e)(k_FL)r^4_s\lambda^2_s(k_Fd)(T/\mu)^2e^{-\mu/T}
\end{equation}
Completely equivalent calculation for the unscreened case, with the
amplitude taken from Eq.~\eqref{A-UC}, gives additional logarithmic
factor for conductance correction $\delta
G\simeq(k_FL)r^4_s\lambda^2_u(k_Fw)(T/\mu)^3\ln^2(\mu/T)e^{-\mu/T}$
as compared to Eq.~\eqref{G}, and similar for the thermopower.

So far we have considered short wires when electrons propagate
ballistically and experience rare backscattering. For longer wires
three-particle collisions become more frequent and simplistic
treatment of the collision integral in perturbation theory by
iterations is not appropriate. For such longer wires electrons reach
the state of \textit{partial} equilibration where backscattering is
achieved by means of multiple collisions at the bottom of the
band.~\cite{TJK,ATJK} While traversing form the right to left Fermi
points electrons perform a random walk in momentum space  with small
momentum step $\delta p\sim T/v_F\ll p_F$ at every collision. For
such diffusive type motion collision integral of the BE can be
reduced to much simpler differential Fokker-Planck form.
Corresponding calculation~\cite{TJK} in this regime shows that
interaction corrections to conductance and thermopower remain
exponentially suppressed, namely $\delta
G=-(e^2/h)(L/\ell_1)e^{-\mu/T}$, where $\ell_1=\sqrt{8\pi mT^3}/B$
and $B\propto r^4_s\lambda^2_s(T/\mu)^5k_Fp_F\mu$ is diffusion
coefficient in momentum space. For even longer wires,
$L\gtrsim\ell_{eq}\propto e^{\mu/T}$, small probability of
scattering at the bottom of the band is compensated by the large
phase space and effects of electronic equilibration in a wire are
nonperturbative. When electrons are fully equilibrated then wire
conductance saturates to the length and interaction strength
independent \textit{universal} value $\delta
G=-(e^2/h)(\pi^2T^2/12\mu^2)$, while thermopower grows from being
exponentially small $\propto e^{-\mu/T}$ to a power law
$S=\pi^2k_BT/6e\mu$, see Refs.~\onlinecite{JTK,TJK,ATJK} for
details.~\cite{Estimates}

\textit{Summary}.-- We have calculated the leading order interaction
corrections to the transport coefficients of a clean single-mode
short 1D quantum wire. Our main results are
Eqs.~\eqref{G}--\eqref{S} for conductance and thermopower. The
dominant scattering mechanism is three-particle collisions which are
not captured by the ideal LL model. Note however that in the
multi-mode case already two-particle inter-channel scattering gives
correction to conductance.~\cite{LFG} We have also emphasized
crucial role of exchange terms in the three-particle amplitude
Eq.~\eqref{A} which was not discussed in the previous
studies.~\cite{Lunde,TJK} Finally, our work also points on the open
questions. First, it is interesting to explore the consequences of
the exchange contributions in the spinfull case. Second, it is of
great interest to understand the fate of interaction corrections in
the limit of strong interactions which must simultaneously coped
with the nonequilibrium effects.

We are sincerely grateful to K. A. Matveev for numerous discussions
which motivated this project. We would like to acknowledge also
useful discussions with L. Glazman and A. Kamenev. This work at ANL
was supported by the U.S. DOE, Office of Science, under contract
DE-AC02-06CH11357.


\begin{thebibliography}{99}
\bibitem{Estimates}
For the numerical estimates one can use $v_F\!\sim\!10^5$m/s,
$k_F\!\sim\!10^8$m$^{-1}$, $\kappa\sim10$ and $w\sim10$nm. This
corresponds to $\mu=\hbar k_Fv_F/2\sim3$meV, $k_Fw\sim1$ and
$e^2/\hbar v_F\kappa\sim1$. For temperatures $T\sim1\div5$K the
interaction-induced correction to conductance of fully equilibrated
wire is $\delta G/G\sim10^{-3} \div10^{-2}$. The effect of
interaction on the thermopower is much more pronounced
$S\sim(k_B/e)\times(0.03\div0.15)$.
\end{thebibliography}
\end{document}